# Classification of Short Segment Pediatric Heart Sounds Based on a Transformer-Based Convolutional Neural Network

Md Hassanuzzaman, Nurul Akhtar Hasan, Mohammad Abdullah Al Mamun, Khawza I Ahmed, Ahsan H Khandoker, *Senior Member, IEEE*, Raqibul Mostafa

***Abstract*—** Congenital anomalies arising as a result of a defect in the structure of the heart and great vessels are known as congenital heart diseases or CHDs. A PCG can provide essential details about the mechanical conduction system of the heart and point out specific patterns linked to different kinds of CHD. This study aims to investigate the minimum signal duration required for the automatic classification of heart sounds. This study also investigated the optimum signal quality assessment indicator (Root Mean Square of Successive Differences) RMSSD and (Zero Crossings Rate) ZCR value. Mel-frequency cepstral coefficients (MFCCs) based feature is used as an input to build a Transformer-Based residual one-dimensional convolutional neural network, which is then used for classifying the heart sound. The study showed that 0.4 is the ideal threshold for getting suitable signals for the RMSSD and ZCR indicators. Moreover, a minimum signal length of 5s is required for effective heart sound classification. It also shows that a shorter signal (3 s heart sound) does not have enough information to categorize heart sounds accurately, and the longer signal (15 s heart sound) may contain more noise. The best accuracy, 93.69%, is obtained for the 5s signal to distinguish the heart sound.

## I. INTRODUCTION

THE human heart is an organ that pumps blood throughout the body by beating nonstop. Heart sounds in the heart reveal essential details about how the heart works. Heart sounds in children are primarily used to monitor and diagnose various cardiac problems. The closing of heart valves and blood passage through the chambers produce cardiac sounds. The significant information of heart sounds the most noticeable and instantly identifiable, often known as S1 and S2. The heart sounds can be shown graphically in a phonocardiogram (PCG) signal. It is achieved using a digital stethoscope or a sensitive microphone to record the heart's oscillations. The PCG signal can visually represent the heart sounds, facilitating in-depth research and interpretation. It helps to diagnose and monitor various cardiac problems by enabling medical professionals to evaluate the timing, intensity, and features of the heart sounds. The PCG signal offers a non-invasive, economical way to assess the heart's function. However, it requires skilled medical professionals.

Children's heart disorders, such as congenital heart disabilities, valve abnormalities, and arrhythmias, can be diagnosed with the use of PCG signal analysis. The PCG signal patterns can be used for future research and treatment strategies. However, there are several obstacles to overcome when evaluating children's heart sounds and PCG signals, such as background noise from breathing noises, ambient noise, and motion artifacts.

Congenital anomalies in the heart's structure are known as congenital heart diseases (CHDs). These disorders can impact the anatomy, function, or both heart and range from minor to severe. Understanding the frequency, death rate, and significance of early identification of CHD is essential for efficient therapy and prevention of this primary worldwide health concern. About 1% of live births worldwide are affected by CHD, the most prevalent birth abnormality. Research estimates that 1.35 million newborns globally are diagnosed with CHD (1). Different demographics and geographical areas have varying rates of CHD prevalence. Research indicates that the prevalence of CHD is more significant in low- and middle-income nations than in high-income nations. For instance, research in the United States found an incidence of 6.9 per

This work has funded by grant (award number: UIU/IAR/02/2019-20/SE/10.) from the Institute for Advanced Research (IAR) of United International University, Bangladesh. The Healthcare Engineering Innovation Center (HEIC), located at Khalifa University in the United Arab Emirates, partially contributed to its funding. Under Award: 8474000132.

Md Hassanuzzaman, Khawza I Ahmed and Raqibul Mostafa is with the Biomedical, IMage and Signals (BIMS), Department of Electrical and Electronic Engineering (EEE), United International University, Dhaka, Bangladesh. (email: hassanuzzaman1503@gmail.com, khawza@eee.uiu.ac.bd, rmostafa@eee.uiu.ac.bd)

Md Hassanuzzaman is also with School of Engineering Mathematics and Technology, University of Bristol, United Kingdom (e-mail: hassanuzzaman1503@gmail.com, qn23845@bristol.ac.uk).

Nurul Akhtar Hasan is with the Pediatric & Congenital Cardiac Program, Children & Women Cardiac Unit, National Heart Foundation Hospital & Research Institute, Bangladesh (e-mail: drhasan.pediicu@gmail.com).

Mohammad Abdullah Al Mamun is with the Division of Neonatal Cardiology, Department of Pediatric Cardiology, Bangladesh Shishu (Children) Hospital and Institute, Bangladesh (e-mail: mamundsh@gmail.com).

Ahsan H. Khandoker is with the Healthcare Engineering Innovation Center (HEIC), Department of Biomedical Engineering, Khalifa University, UAE (e-mail: ahsan.khandoker@ku.ac.ae).

1,000 live births (2), but a study in India found 8.2 per 1,000 live births (3). Numerous variables might be shortcomings for these geographical discrepancies, including genetic predisposition, environmental factors, prompt treatments, and disparities in healthcare resources and access. These variables also affect the death rate linked to CHD. About 25% of all baby fatalities worldwide are attributable to congenital abnormalities, with CHD being the primary cause of death (1). Some severe types of CHD can result in death during the first few days or weeks of life, with the most significant fatality rate occurring during this time. Higher death rates may also be attributed to the long-term consequences of CHD. Adults with CHD are more likely to experience problems such as arrhythmias, infective endocarditis, and heart failure, all of which can shorten their survival (4).

TABLE I
COMPARISON OF DIFFERENT STUDIES

| Author (Year) | Dataset | Duration | Collection Area | Age (Year) | Classifier | Accuracy% |
|---|---|---|---|---|---|---|
| Hettiarachchi (2021) (26) | PhysioNet2016 | 3.5s | Pulmonary valve, Aortic valve, Mitral valve, and Tricuspid | Unknown | CNN | 90.41 |
| Chen (2022) (27) | PhysioNet2016 | 5s | Pulmonary valve, Aortic valve, Mitral valve, and Tricuspid | Unknown | CNN+LSTM | 86 |
| Potes (2016) (28) | PhysioNet2016 | 2.5s | Pulmonary valve, Aortic valve, Mitral valve, and Tricuspid | Unknown | Ensemble CNN | 89 |
| Aziz (2020) (29) | Self | 5s | Between the third and fourth left intercostal space | | SVM | 95.24 |
| Gharehbaghi (2020) (30) | Self | 10s | APEX | 1 to 15 | CNN | 88.40 |
| Lv (2021) (31) | Self | 10s | Pulmonary valve, Aortic valve, Mitral valve, and Tricuspid | 2.4 to 3.1 | CNN | 96 |
| Bozkurt (2018) (32) | Self | 4-10s | Apical, Lower Left, Upper Left, Right parasternal | Unknown | KNN | 81.50 |
| Sepehri (2018) (33) | Self | 10s | Unknown | 1 to 18 | SVM | 87.45 |
| Gharehbaghi (2017) (34) | Self | 10s | APEX | 1 to 12 | SVM | 86.40 |
| This work | Self | 15s,5s,3s | Pulmonary valve, Aortic valve, Mitral valve, and Tricuspid | 5 months to 18 | Transformer based CNN | 93.67,93.69, 92.41 |

For several reasons, it is essential to diagnose CHD early. First, it makes suitable management techniques and prompt actions possible. The prognosis for newborns and kids with congestive heart failure has dramatically improved because of developments in medical technology and surgical methods. Better long-term results may be achieved by using these treatment choices efficiently, which is ensured by early diagnosis. Early detection and treatment of children with severe CHD considerably increased survival rates and decreased the need for emergency treatments, according to research published in the Journal of the American College of Cardiology (5). Second, the prevention of CHD-related problems is made possible by early diagnosis. The development of heart failure, pulmonary hypertension, and other problems that may result from untreated or inadequately managed congestive heart failure can be avoided or minimized with prompt intervention. Finally, for those with CHD, an early diagnosis improves their overall quality of life. Early therapies can boost physical and cognitive development, lessen symptoms, and improve heart function. Additionally, it enables families to receive the proper therapy and assistance, assisting them in overcoming the difficulties related to CHD.

Heart auscultation is a problematic clinical procedure, especially in younger patients. Apart from requiring much skill and experience, there are difficulties with perception: In many cases, especially in pediatric cardiac auscultation, the heart murmurs and sounds are barely audible and are primarily low-frequency components (carrying discriminative characteristics for detecting abnormalities). They are also often accompanied by high noise (breath, ambient noise, scratches from microphone movement). Healthcare practitioners can access various diagnostic instruments and methods, including PCG and auscultation, to identify CHD in newborns, children, and adults. A few of the current methods for detecting CHD include echocardiography, electrocardiography (ECG), magnetic resonance imaging (MRI), and computed tomography (CT) scanning.

Ultrasound waves are used in echocardiography to provide real-time pictures of the anatomy and function of the heart. In-depth details on the heart's chambers, valves, blood flow patterns, and structural anomalies may be obtained using echocardiography (6). Regarding CHD, an ECG can provide essential details about the heart's electrical conduction system and point out specific patterns linked to different condition forms. ECG is frequently used with other diagnostic instruments since it is insufficient to diagnose structural cardiac problems. Cardiac MRI may offer detailed information on the architecture and function of the heart, and it is beneficial in evaluating complicated instances of congenital heart disease. However, cardiac MRI is costly, time-consuming, and needs specific knowledge and equipment from medical professionals. It is usually saved for situations in which the results of other imaging modalities have been equivocal or when comprehensive anatomical information is required (7). A CT scan is an additional imaging modality that is used to diagnose congestive heart failure. It creates cross-sectional pictures of the heart and blood arteries using computer processing. When assessing complicated instances of CHD, aortic abnormalities, and coronary artery anomalies, CT scans are beneficial. Ionizing radiation exposure is a risk associated with CT scans, particularly for young patients. As a result, the advantages of using CT scans to diagnose CHD must exceed any possible dangers (7).

Research on automatic cardiac anomaly detection using PCG signals is extensive. Hasan (8) discussed the identification of early cardiovascular abnormalities using PCG. Swarup (9) examines how the stethoscope has changed and emphasizes the advancements brought forth by the contemporary digital stethoscope. This study also highlights the promise of digital stethoscope technology in telemedicine since it enables distant evaluation and enhances access to healthcare in underserved areas. Techniques for acquiring and preprocessing cardiac sounds are essential for diagnosing CHD. Ghosh (10) thoroughly analyses the most recent methods for acquiring and preprocessing cardiac sounds. It shows how to segment the first and second heart sounds (S1, S2), denoise the PCG signal, extract features from the separated heart sound components, and use feature selection algorithms. In the summary and organization of the research, Burns (11) highlights the sensitivity and accuracy of intelligent PCG in identifying pathologic murmurs and particular lesions, such as aortic stenosis and regurgitation, and its use in diagnosing CHD in pediatric patients. Liu et al. (12) and Marascio and Modesti (13) provided thorough analyses of the developments in feature selection and automated classification techniques.

Segmentation, feature extraction, machine learning, and data are crucial components of an automatic classification system, just like they are for several other automated classification problems. All these components (and how they interact) have been extensively documented in the several hundred papers that comprise this massive body of work. Since feature extraction entails converting unprocessed data into a format that machine learning algorithms can use efficiently, it is a crucial stage in automatic categorization systems. On the other hand, new developments in deep learning methods have demonstrated encouraging outcomes in automatically extracting pertinent characteristics from the data. Features for automated PCG classification may be categorized into time, frequency, statistical, and time-frequency domain characteristics (13). The following are typically included in the feature vector of research using time-domain features: duration measures (for S1, S2, diastole, systole, R-to-R) and their ratios (e.g., the ratio of the systolic interval to the heartbeat), as well as other standard time-domain features like the zero-crossing rate. An open-source system with these features is offered (12) and is the basis for the PhysioNet-2016 challenge (14). Spectral (frequency domain) features cover various spectrum representations and metrics, similar to automated sound categorization tasks. The low-frequency bands include essential information that may be successfully simulated for generating discriminative features for PCG signals, according to Schmidt et al.'s (15) examination of various spectral characteristics for automatic PCG classification. These characteristics included power in octave bands, instantaneous frequency and amplitude (IFA), and parametric models for the spectra. By utilizing the Hilbert-Huang transform to break down PCG signals into intrinsic mode functions (IMFs), Arslan and Karhan (16) provide a novel, efficient, and stable feature generator. The many machine learning techniques applied to classification include clustering (17), support vector machines (18), decision trees (19), KNN (20), ensemble (21), and artificial neural networks (ANN) (22). So far, it has been shown that machine learning-based automated heart sound classification methods depend on complex characteristics. These characteristics are often calculated by hand. Because the extracted features are very complex, subjective bias and variations can be observed even with the high output accuracy. Krishnan et al. (23) proposed a deep learning method that eliminates the requirement to first segment PCG data into fundamental heart sounds. A few deep learning-based computer-aided approaches for pediatric CHD diagnosis were developed using 15s heart sounds (24, 25). Table I provides a comparative overview of the literature regarding the dataset, the age range of samples collected, signal duration methods of classification, and evaluation parameters of the classifier used in the experiment (26 – 34).

This study establishes a large dataset of children's heart sounds using a digital stethoscope from 751 distinct pediatric patients. Next, a trial is conducted to get optimal precision by altering the PCG signal's duration. A signal quality evaluation is also used in the experiment to exclude unsuitable signals for further examination. Next, using MFCCs-based feature taking as an input, a transformer-based residual 1D CNN deep learning model is developed. The suggested model is also assessed by changing the signal quality assessment indicators to determine the minimal heart sound signal duration. To determine the relationship between the model's performance and demographic data, the suggested model is assessed using patient demographic information, such as age and gender.

TABLE II
GENDER DISTRIBUTION OF COLLECTED DATA.

|  | CHD, N (%) | Non-CHD, N (%) | All, N (%) | pValue |
|---|---|---|---|---|
| Male | 257 (56.36) | 190 (64.41) | 447 (59.52) | **0.028** |
| Female | 199 (43.64) | 105 (35.59) | 304 (40.48) |  |

TABLE III
AGE DISTRIBUTION OF COLLECTED DATA.

| Group | Age (Years) | CHD, N (%) | Non-CHD, N (%) | All, N (%) | pValue |
|---|---|---|---|---|---|
| Infancy | 1 month to 2 | 125 (27.41) | 137 (46.44) | 262 (34.89) | **<0.001** |
| Childhood | 2 to 12 | 306 (67.11) | 152 (51.53) | 458 (60.98) |  |
| Adolescence | 12 to 16 | 22 (4.82) | 6 (2.03) | 28 (3.73) |  |
| Adult | > 16 | 3 (0.66) | 0 | 3 (0.4) |  |
| Total |  | 456 (60.71) | 295 (39.28) | 751 |  |

TABLE IV
ANTHROPOMETRY OF COLLECTED DATA.

|  | Weight (kg) | | Height (cm) | | BMI | |
|---|---|---|---|---|---|---|
|  | CHD | Non-CHD | CHD | Non-CHD | CHD | Non-CHD |
| Mean | 15.98 | 15.18 | 103.93 | 98.4 | 14.02 | 14.61 |
| STD | 8.61 | 10.28 | 24.59 | 26.95 | 2.38 | 3.42 |
| Max | 62.5 | 80 | 242 | 177 | 27.34 | 29.6 |
| Min | 3.4 | 2.3 | 58 | 54 | 4.54 | 4.7 |
| pValue | **<0.001** | | 0.004 | | 0.088 | |

## II. DATA COLLECTION PROCEDURE

The inability of pediatric patients to completely follow the directions given by the physicians throughout the collecting procedure makes it challenging to record and categorize pediatric heart sounds. The children's cries, screams, coughing, chatting, movement, and breathing noises add to the noise level during the heart sound collection. The main conclusion is that the lack of publicly available statistics prevents academics from conducting more research on pediatric CHD. Regarding sound pressure, heart rate, pathology, and heart sound components, adult heart sounds also vary from pediatric heart sounds. For this reason, they are inappropriate for direct algorithmic experiments and pediatric cardiac sound analysis.

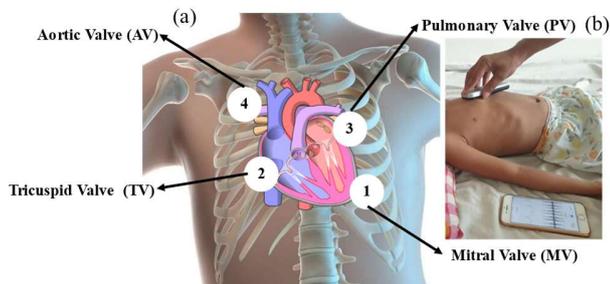

Fig. 1. The auscultation areas from where the PCG signal is acquired (a) schematic diagram of four auscultation areas: mitral valve (MV), tricuspid valve (TV), pulmonary valve (PV), and aortic valve (AV) and (b) Collect PCG signal using EKO device.

A digital stethoscope is employed to verify the minimum signal duration required for accurate classification and diagnosis by implementing a proposed deep learning method.

The PCG signals in this experiment were obtained using the Eko DUO ECG + Digital Stethoscope, the first combined EKG stethoscope to get FDA authorization (35), shown in Figure 1. The DUO can operate continuously for up to 10 hours after a full charge. Data was gathered from healthy individuals (non-CHD) and those with CHD in clinical settings. The project received ethical approval from the regional medical ethics committees of Bangladesh Shishu (Children) Hospital & Institute's (Ethics Approval Number: Admin/1714/BSHI/2021) and National Heart Foundation Hospital & Research Institute's (Ethics Approval Number: NHFH & RI 4-14/7/AD/1132). Before the data is recorded, the persons or their legal guardians provide informed consent for collecting and using it. Assent is requested in addition to informed consent (where appropriate) for minors under fifteen. Every participant provided information about their medical history in the consent form. Each patient had four or more PCG signals recorded consecutively at each of the four auscultation areas: the mitral valve (MV), tricuspid valve (TV), pulmonary valve (PV), and aortic valve (AV). All positions of CHD and non-CHD signals are shown in Figures 2 and 3. Each signal is captured for 15 seconds at a 4000Hz sample frequency that contains 60000 sampling points. The datasets contained the PCG signals from 751 people (total signals: 3435). Regarding demographics, the collected data is slightly skewed towards the male population (59.52%), shown in Table II, with most patients ranging from infancy to adolescence (5 months to 16 years), as shown in Table III. Table IV represents the patient's anthropometry. The statistical analysis of all available patient data is conducted using the one-way student t-test analysis of variance (ANOVA).

Statistical significance is indicated if the p-value is less than or equal to 0.05. The pediatric cardiologist reviewed the signals gathered following hospital guidelines. Echocardiography, 12-lead ECG, chest x-ray, CT scan, and clinical verifications such as auscultation of the heart sound are among the inclusion criteria. The exclusion criteria included being extremely obese and having a lot of background noise, such as snoring, humming, and crying.

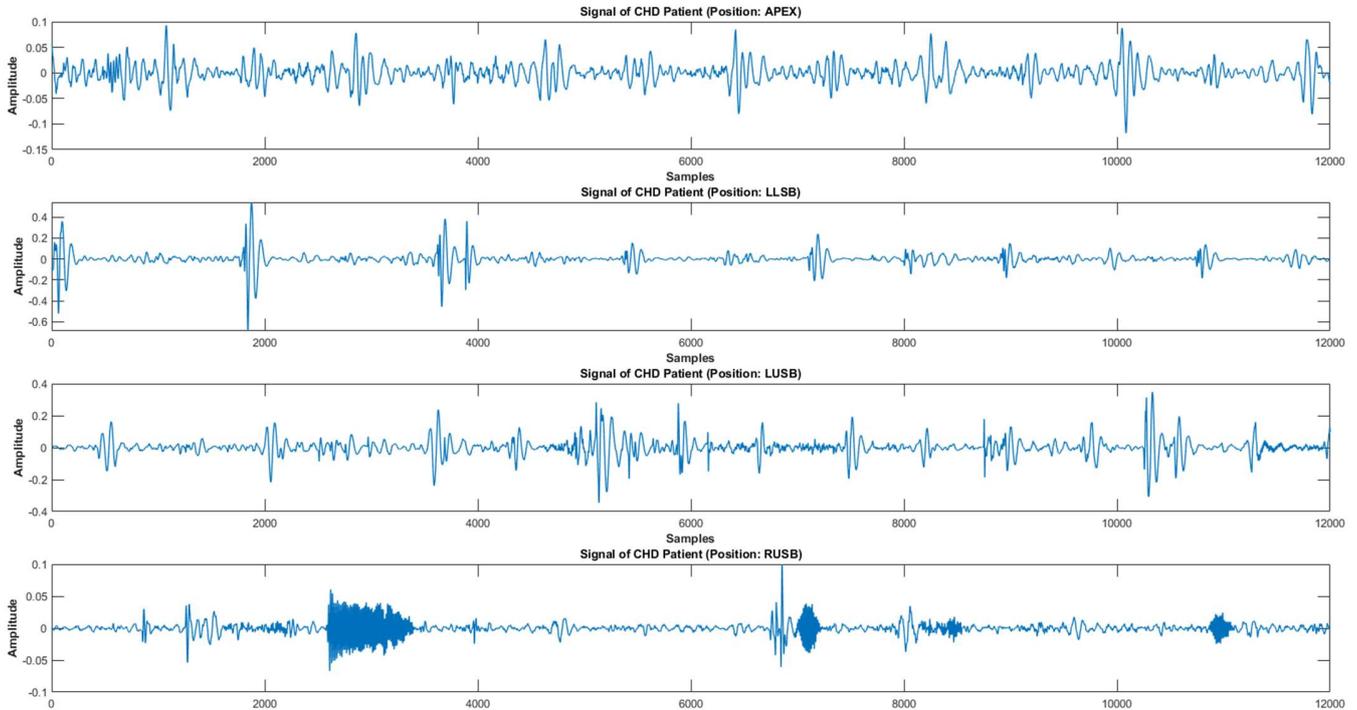

Fig. 2. The signals of CHD of four positions: mitral valve (MV), tricuspid valve (TV), pulmonary valve (PV), and aortic valve (AV) (from top to bottom).

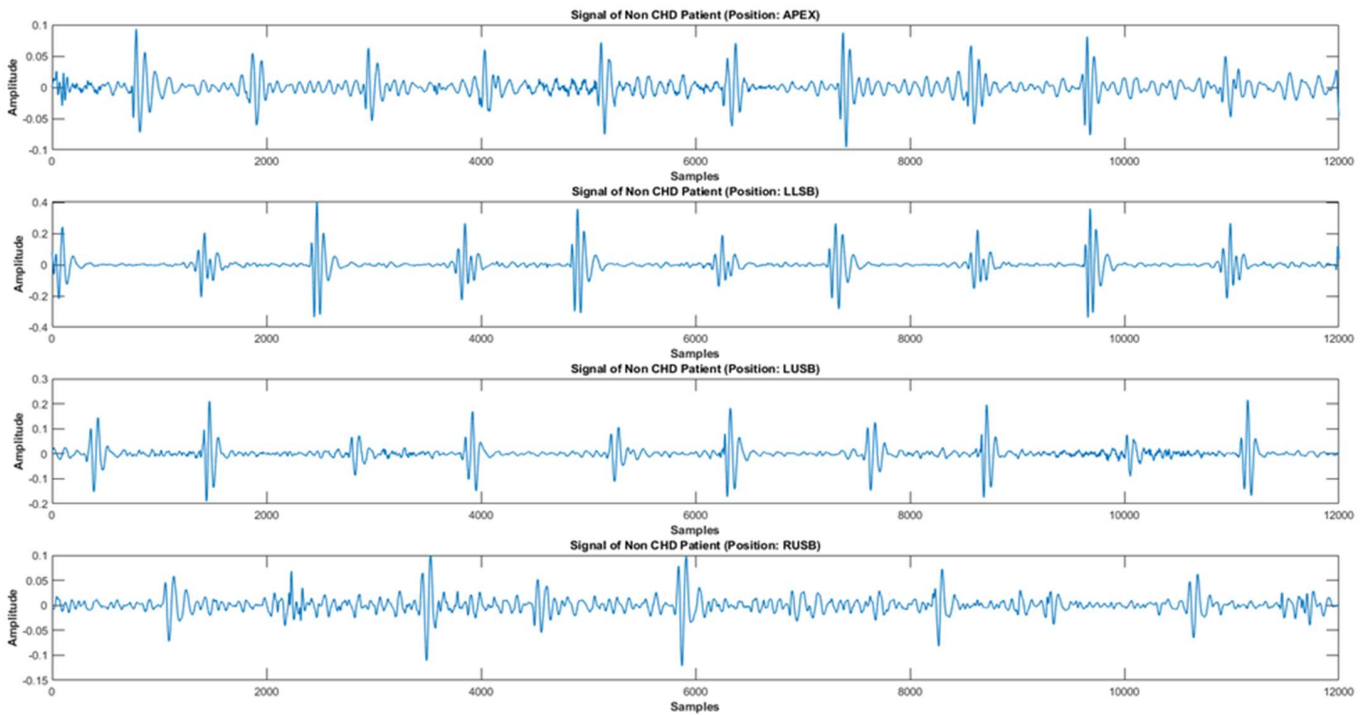

Fig. 3. The signals of non-CHD of four positions mitral valve (MV), tricuspid valve (TV), pulmonary valve (PV), and aortic valve (AV) (from top to bottom)

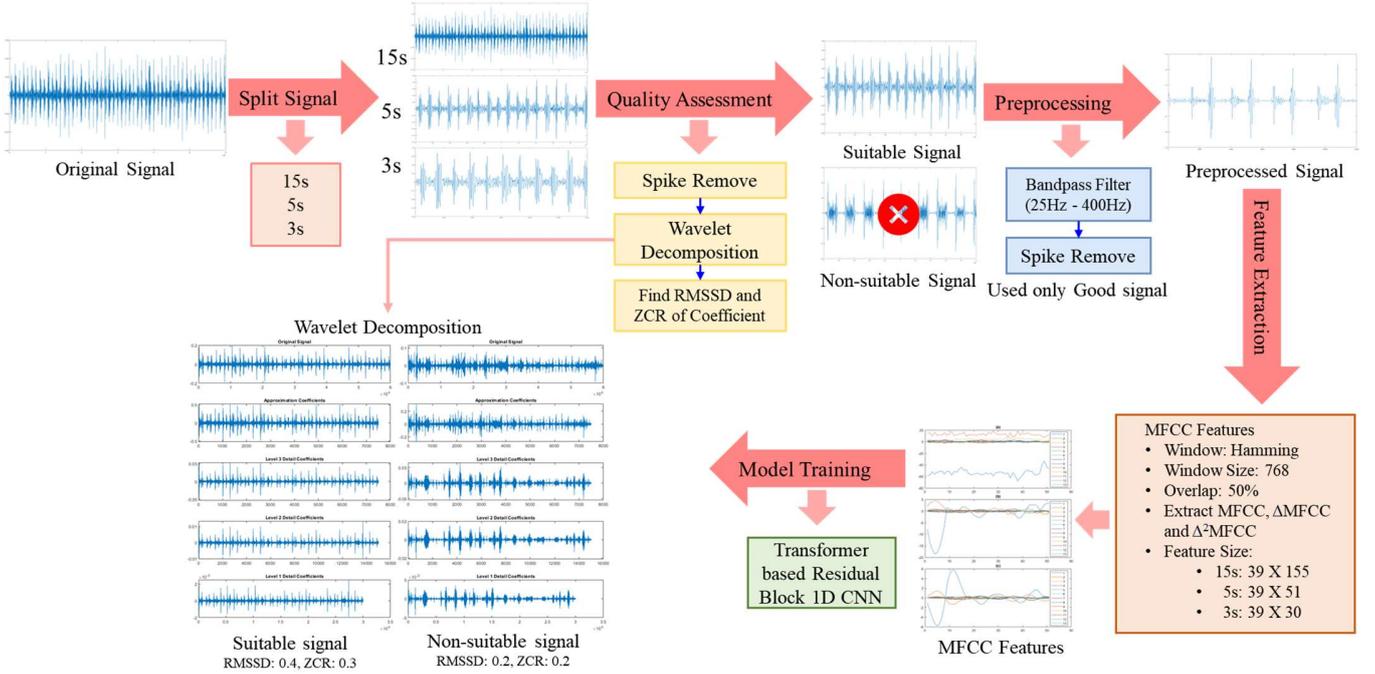

Fig. 4. The system architecture of the proposed method to verify the heart sound classification accuracy by varying signal duration. It includes signal split, quality assessment, preprocessing, feature extraction, and training of the proposed model.

## III. METHODS

The architecture of the proposed CHD auscultation algorithm is shown in Figure 4. The proposed algorithm includes signal quality assessment, preprocessing, MFCC feature extraction, and train transformer-based residual 1D CNN model training. The experiment's main objective is verifying the accuracy of the heart sound classification by varying signal duration. At first, the signal is split into 5s and 3s PCG signals containing 20000 and 12000 sampling points, respectively. The sampling point of the 15s signals is 60000.

### A. Quality Assessment

The quality evaluation method is used to evaluate the signal quality because the heart sounds were recorded in actual environments, which may have noises including the surrounding environment and conversations, as well as bronchial and digestive sounds. Less noise also complies with the clinical diagnostic criteria; thus, the heart sound's quality must be evaluated before using the signal. This evaluation helps ensure accurate diagnosis and prevents potential misinterpretation of the heart sound signal.

First, a three-level discrete wavelet transform is applied to the Daubechies wavelet to decompose heart sound data. An approximation coefficient is obtained at the three wavelet decomposition levels. It is applied in determining the assessment criteria (36). As discussed below, two metrics are used to evaluate the suitability of the PCG signal: the RMSSD and the ZCR of the three-level wavelet decomposition approximation coefficients. Figure 5 represents the 3-level wavelet coefficient and approximate coefficient of suitable and non-suitable signals.

RMSSD: The successive differences within the signal are determined to compute the RMSSD. Their root mean square is computed using equation (1). This metric effectively captures the fluctuations in the signal. To ensure the signal's suitability for processing, the RMSSD should be under the specified threshold.

$$RMSSD = \sqrt{\frac{\sum_{i=1}^{N-1}(x_{i+1} - x_i)^2}{N - 1}} \quad (1)$$

In the equation, $x$ denotes the approximate coefficients of a three-level wavelet decomposition sequence, and $N$ is the sequence length.

ZCR: Equations (2) and (3) are used to calculate the ratio of zero crossings, or intersections with the x-axis, to the signal length, which is used to assess the adequacy of the signal. This ratio, which captures fluctuations linked to heartbeats, is called the ZCR and shows how frequently the signal's sign changes (25). When there is much high-frequency noise in the signal, the value of ZCR becomes higher.

$$RZC = \frac{\sum_{i=1}^{N-1}[sign(x_i \times x_{i+1}) \cap |x_{i+1} - x_i| > 0]}{N - 1} \quad (2)$$

$$sign(x) = \begin{cases} 1, & x \leq 0 \\ 0, & otherwise \end{cases} \quad (3)$$

### B. Preprocessing

Noise Removal: The preprocessing of heart sound signals is the foundation of the whole classification method. A fifth-order Butterworth bandpass filter with a 25–400 Hz cut-off frequency is employed to eliminate high-frequency interference, baseline wandering, and low-frequency aberrations from the initial heart sound signals. The heart sound information's primary distribution frequency band is preserved after the filter.

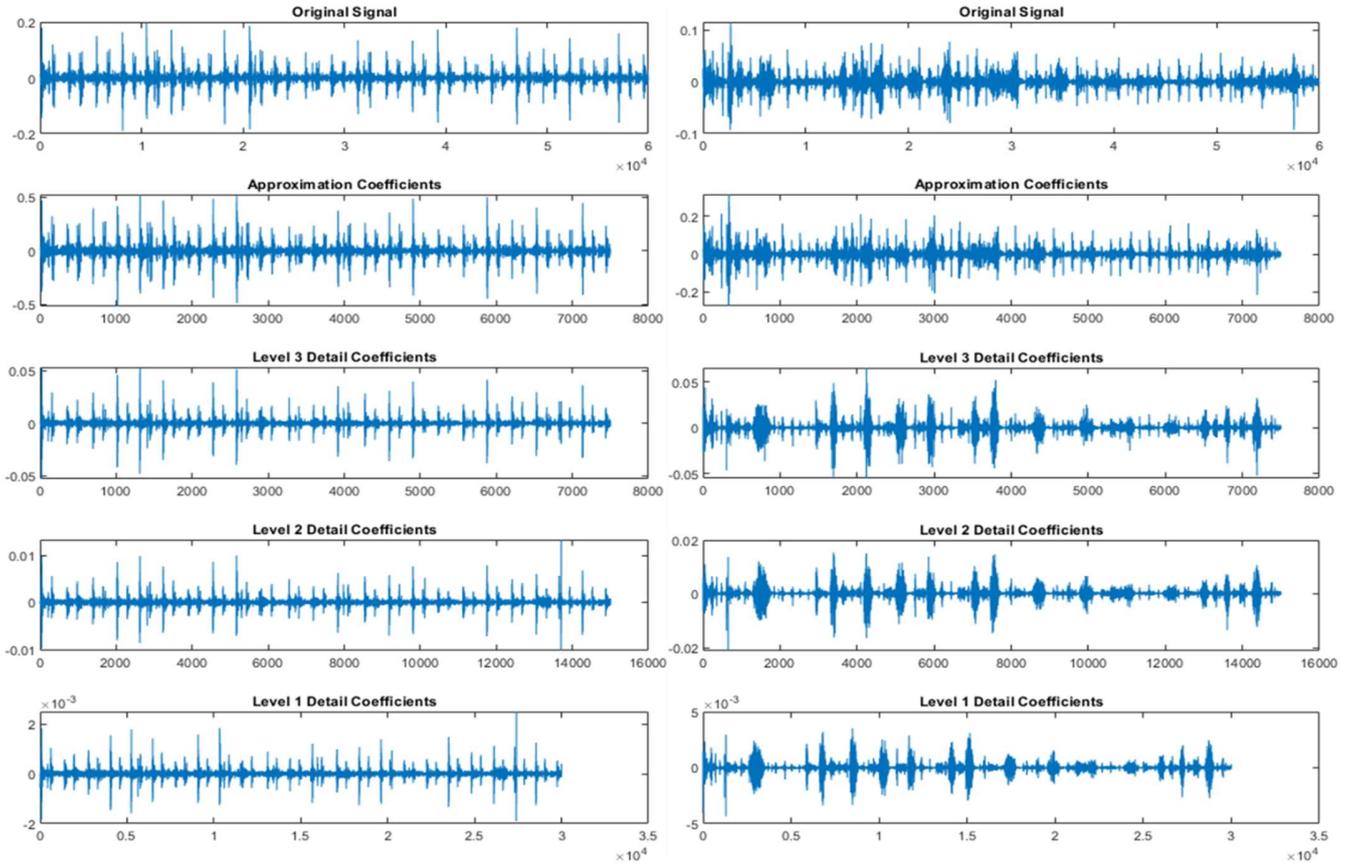

Fig. 5. Discrete wavelet transformations (DWTs) are used in the picture to decompose a signal into its detailed components. The panels below display the decomposed components, while the top panel shows the original signal. (Left) an appropriate signal with an RMSSD of 0.4 and a ZCR of 0.3; (Right) an unsuitable signal with an RMSSD of 0.2 and a ZCR of 0.2.

Spike Removal: The Schmidt algorithm (37) is used to locate and eliminate spikes brought up by the acquisition procedure. An iterative process is used to identify and remove the spikes to address the frequency of friction spikes larger in amplitude than heart sounds. A discrete 500 ms window segmentation process is used to extract the maximum absolute amplitude (MAA) from the recorded data. A crucial benchmark is established, requiring further measures if a minimum of one MAA surpassed three times the median value of all MAAs. In these cases, a set of defined actions occurred, starting with the window chosen to have the maximum MAA. The peak of the detected noise spike inside this window is then determined, and the zero-crossing points of the noise spike are established at the last zero-crossing point before the MAA and the first zero-crossing point following the amplitude maximum, respectively. The process resumed after the selected noise spike is methodically replaced with zero data. This thorough procedure ensured that friction spikes are effectively reduced and ended when all specified procedures are finished. Although signal preparation is the primary emphasis of an autonomous heart sound analysis system, preprocessing methods are not addressed in this study because they are generally well-known. In particular, the effect of signal duration variation on the accuracy of automated heart sound classification is discussed.

MFCC feature extraction: MFCC is utilized to derive PCG signal features for the frequency domain in categorizing heart sounds. Thirteen one-dimensional MFCCs, thirteen one-dimensional ΔMFCCs, and thirteen one-dimensional $\Delta^2$MFCCs are retrieved in this experiment; thirty-nine one-dimensional useful features are collected from each signal. Later, a transformer-based residual 1D CNN model is trained to classify heart sounds. MFCC is explained in detail below. Figure 6 represents the one-dimensional MFCC, ΔMFCC, and $\Delta^2$MFCC features of non-CHD and CHD heart sounds.

The non-linear relationship between the frequency of a sound and the human ear is reflected in the Mel-frequency cepstrum. This relationship can be expressed in equation (4):

$$mel(f) = 2595 * \left(1 + \frac{f}{700}\right) \quad (4)$$

The MFCC algorithm consists of the following steps:

1) **Pre-emphasis:** In this stage, the signal's high-frequency components are amplified while its low-frequency components are diminished. The formula of the pre-emphasis filter is:

$$S_{pre-emphasis}(n) = s(n) - \alpha \cdot s(n-1) \quad (5)$$

where $s(n)$ is the input signal, $S_{pre-emphasis}(n)$ is the output signal, and $\alpha$ is the pre-emphasis coefficient.

2) **Window framing:** The heart sound signal is non-stationary, yet it achieves a steady condition between 24 and 256

milliseconds. The experiment creates a short frame using 192 ms following equation (6), and feature extraction is carried out within a short frame. 50% overlap is used between consecutive frames to create a smooth transition.

$$S_{framed}(m, n) = s_{pre-emphasis}(n - m \cdot N_w) \quad (6)$$

where $N_w$ is the window length.

The window function is then used on every frame to reduce spectral leakage. Since the Hamming window function can prevent the leaking phenomena, it is used for signal framing in the experiment (38). It is expressed in equation (7).

$$W(n) = (1 - \alpha) - \alpha \cdot cos\left(\frac{2\pi n}{N-1}\right), 0 \leq n \leq N - 1 \quad (7)$$

where $\alpha$ is empirically determined to be 0.46 (39), and $N$ is the number of samples for each frame.

3) **Discrete Fourier Transform (DFT):** DFT is used to convert the time-domain heart sound signal, $x(n)$, into a frequency domain signal to obtain a spectrum $X(k)$ following the equation (8):

$$X(k) = \sum_{n=0}^{N-1} x(n) \cdot e^{-j\frac{2\pi}{N}kn} \quad (8)$$

where $X(k)$ is the $k$th frequency component of the DFT, $x(n)$ is the $n$-th input data point, $N$ is the total number of data points, $j$ is the imaginary unit.

4) **Power spectrum calculation:** By utilizing the signal spectrum $X(k)$ as the square of its modulus, one can obtain the power spectrum $S(k)$ using the subsequent equation (9):

$$S(k) = \frac{1}{N}|X(k)|^2 \quad (9)$$

5) **Mel Filter bank:** The power spectrum S(k) is passed through a set of mel-scale triangular filter banks to mimic the non-linear human ear perception of frequency and obtain a mel spectrum. The product of P(k) and filters $H_m(k)$ is calculated at each frequency. If we define a triangular filter bank with M filters, the frequency response of the triangular filter $H_m(k)$ is calculated as follows in equation (10).

$$H_m(k) = \begin{cases} 0, & k < f(m-1) \\ \frac{k - f(m-1)}{f(m) - f(m-1)}, & f(m-1) \leq k \leq f(m) \\ \frac{f(m+1) - k}{f(m+1) - f(m)}, & f(m) \leq k \leq f(m+1) \\ 0, & k > f(m+1) \end{cases} \quad (10)$$

where f(m) is the mel triangular filter's center frequency. It is expressed as

$$\sum_{m=0}^{M-1} H_m(k) = 1 \quad (11)$$

6) **Log Transformation:** The logarithm energy spectrum $S_{mel}(m)$ at each frame is then obtained by applying a logarithmic operation, shown in equation (12), to simulate human loudness perception.

$$S_{mel}(m) = ln\left(\sum_{k=0}^{N-1} S(k) \cdot H_m(k)\right), 0 \leq m \leq M \quad (12)$$

where $S(k)$ is the power spectrum and $H_m(k)$ is the filter bank, and $M$ is the number of filter banks.

7) **Discrete Cosine Transform (DCT):** To decorrelate the mel-frequency cepstral coefficients $MFCC_i$, the above logarithmic spectrum is subjected to the DCT.

$$MFCC_i = \sum_{i=0}^{N-1} S_{mel}(m) \cdot cos\, cos\left(\frac{\pi n(m - 0.5)}{M}\right),$$

$$n = 1,2, \ldots\ldots\ldots, \quad L \quad (13)$$

where L is the order of the MFCC coefficient, and M denotes the number of filter banks.

8) **ΔMFCC and Δ²MFCC feature:** Given the previous explanation, the MFCC coefficients that are computed only captured the static aspects of the heart sound signal. The dynamic information of the heart sound spectrum also provides a wealth of information, which may be utilized to increase the classification accuracy further because the human ear is more sensitive to the dynamic feature of an acoustic signal. The first and second differences of MFCC are used to derive the ΔMFCC and Δ²MFCC, which reflect the dynamic information of the heart sound signal. The following formula is used to get coefficients of ΔMFCC and Δ²MFCC.

$$\Delta MFCC_i = \frac{\sum_{n=1}^{N} n \times (MFCC_{i+n} - MFCC_{i-n})}{2\sum_{n=1}^{N} n^2} \quad (14)$$

$$\Delta^2 MFCC_i = \frac{\sum_{n=1}^{N} n \times (\Delta MFCC_{i+n} - \Delta MFCC_{i-n})}{2\sum_{n=1}^{N} n^2} \quad (15)$$

In equations (14 and 15), $\Delta MFCC_i$ is the ΔMFCC at frame i, $\Delta^2 MFCC_i$ is the Δ²MFCC at frame i, $MFCC_i$ is the MFCC at frame i, and N is the number of frames to consider for the computation.

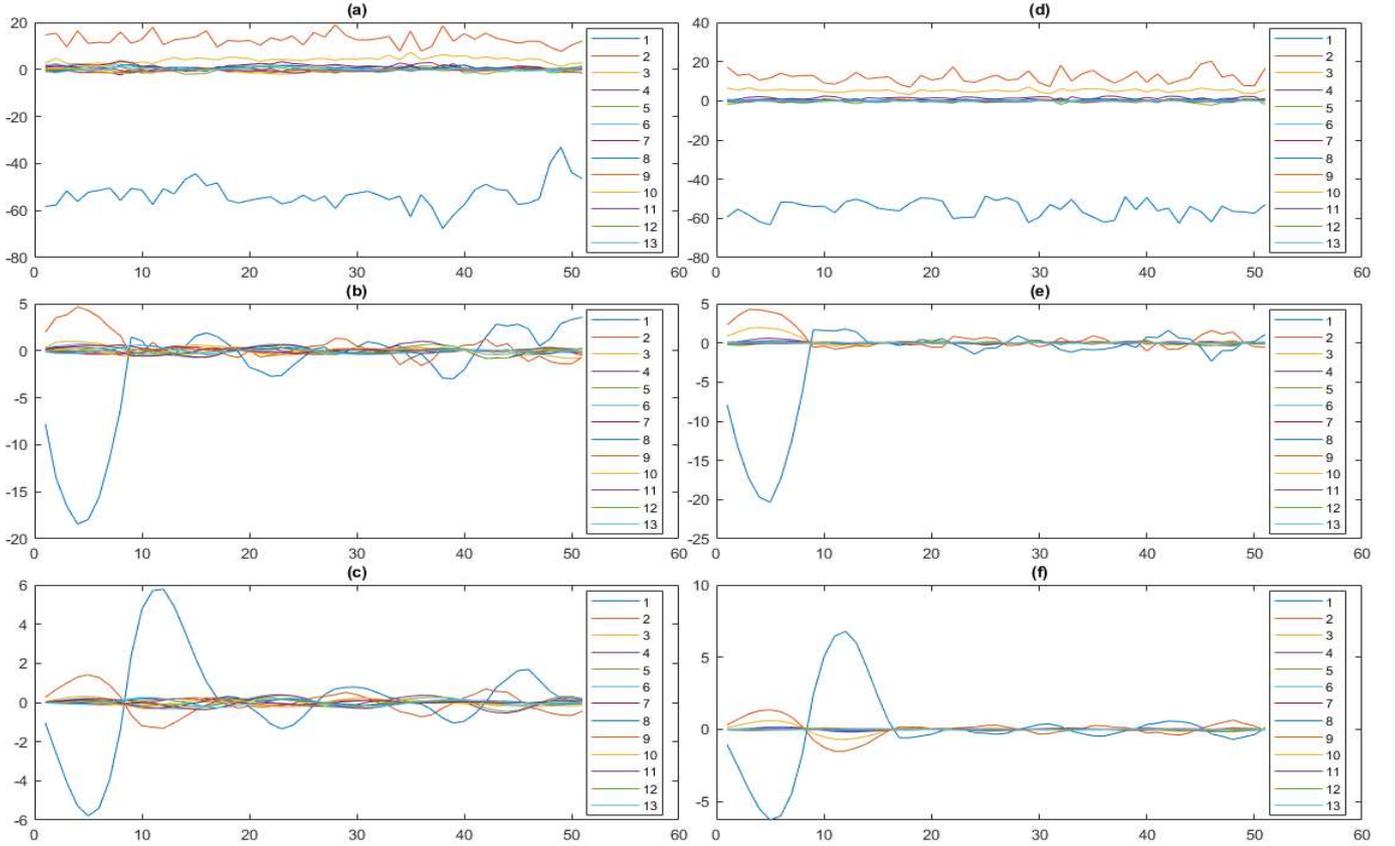

Fig. 6. It represents all extracted MFCC feature curves of 5s heart sound: (a) MFCC features of non-CHD heart sound, (b) ΔMFCC features of non-CHD heart sound, (c) Δ²MFCC features of non-CHD heart sound, (d) MFCC features of CHD heart sound, (e) ΔMFCC features of CHD heart sound, and (f) Δ²MFCC features of CHD heart sound.

### C. CHD Detection Model

In order to improve the performance in classifying the heart sound data, this study offers a unique deep learning architecture that combines the advantages of transformers and residual 1D convolutional neural networks (CNN), as shown in Figure 7. A residual 1D CNN model that is transformer-based and designed, especially for non-stationary cardiac sound signals, is provided. The design has two main parts: the transformer encoder and residual 1D CNN layers to extract effective features from heart sounds. The remaining 1D CNN layers concentrate on collecting local patterns and features, and the transformer encoder extracts long-range relationships in the sequential input. With residual connections added for improved training stability and speed, this hybrid architecture combines transformers' global dependency modeling power with the local feature extraction expertise of 1D CNNs. The system architecture is explained in detail below.

1) **Input:** Each signal has extracted MFCC characteristics utilized as an input. In the experiment, the features' sizes are 39 X 155 for the 15-second signal, 39 X 51 for the 5-second signal, and 39 X 30 for the 3-second signal.
2) **Feature Encoder:** Local features and patterns are extracted from the heart sound signal using 1D convolutional layers with a kernel size 3. These layers establish the foundation for additional analysis by capturing close links within the sequence. Batch normalization (BN) and a rectified linear unit (ReLU) are activation layers after the 1D convolution layer. A convolutional mathematical expression is defined in equation (16). By normalizing intermediate outputs during training, batch normalization (BN) successfully tackles the internal covariate shift problem. It leads to faster convergence, more robust learning, and increased network stability. It also aids in lessening overfitting and regularizing the model. ReLU adds non-linearity to the network by setting negative values to zero and permitting positive values to flow through. The model's innate non-linearity makes discovering intricate links in the data more accessible. Additionally, ReLU is used to tackle the activation functions-related vanishing gradient issue.

$$C_i^{lj} = a\left(b_j + \sum_{m=1}^{M} w_m^j x_{i+m-1}^j\right) \quad (16)$$

Where $x_i = [x_1, x_2, \ldots x_n]$ is the input, n is the total number of points, l is the layer index, a is the activation function, b is the bias of the $j^{th}$ feature map, M is the kernel size, $w_m^j$ is the weight of the $j^{th}$ feature map and $m_{th}$ filter index.

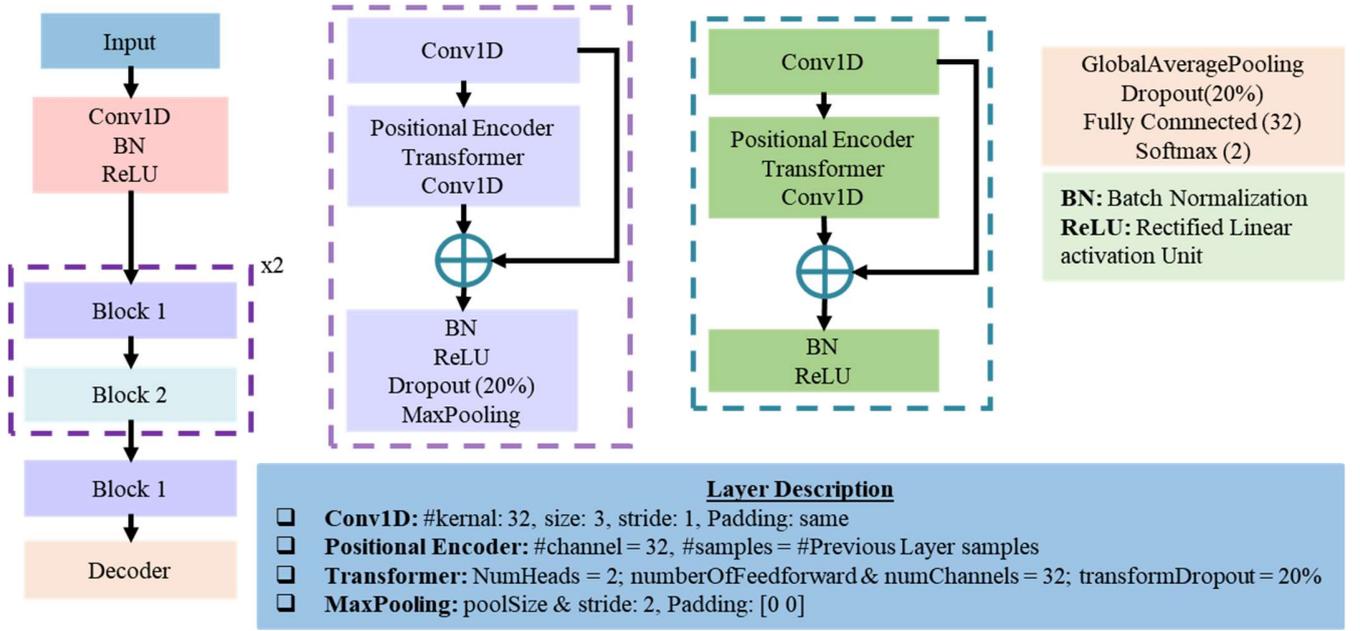

Fig. 7. The system architecture of transformer-based residual 1D CNN deep learning model.

1) **Block 1:** Block -1 process uses a Conv1D, positional encoder (PE), transformer, BN, ReLU, dropout, and max pooling layer. It also has a residual connection with the Conv1D layer and BN layer. The parameter of the Conv1D layer in block-1 is the same as before used Conv1D.

a. **Positional Encoder:** This experiment's positional encoder layer becomes crucial in integrating temporal information. This layer gives each component of a sequence a distinct position or timestamp, which helps the model understand the temporal correlations in the signal. It is essential because it allows the model to classify heart sounds accurately based on the timing, length, and dynamic nature of individual occurrences. Simultaneously, positional encoding guarantees that every feature vector in a sequence is assigned a unique representation, a procedure essential to temporal sequence analysis, especially before the implementation of self-attention transformer layers. In the study, the set of sinusoidal signals within the PE is selected as the unique representation defined in equations (17 and 18). The formula that determines the frequency of these signals considers the feature vector's location, which improves the model's ability to comprehend the temporal understanding of feature extraction and allows the self-attention transformer layers to focus on distinct sequence points selectively. The network's capacity to identify the order and timing of auditory events is strengthened by this method. The effectiveness of suggested deep learning models in heart sound classification is increased due to the notable improvements in contextual awareness and precision of classifications linked to different cardiac diseases.

$$PE_{(pos,2i+1)} = \sin\left(\frac{pos}{\omega^{\frac{2i}{d_{model}}}}\right) \quad (17)$$

$$PE_{(pos,2i+1)} = \cos\left(\frac{pos}{\omega^{\frac{2i}{d_{model}}}}\right) \quad (18)$$

Where $pos$ and $i$ represent the position in the feature and time-sequence dimensions, respectively, $d_{model}$ defines the dimension of the model. The $\omega$ is the encoding value. In the study, $\omega$ is selected as low as 10,000 to make sure a distinct encoding at different locations.

The encoding approach can capture complex periodic patterns in the input data using the features of sine and cosine functions. These functions are ideal for this purpose since they are good at depicting cyclical fluctuations with discrete phases. A more subtle distinction between elements is made possible by selecting 10,000 elements. This granularity allows the encoder to encode finer changes and correlations within the input sequence by offering a more comprehensive range of positional information.

b. **Transformer:** Applications of the transformer layer to signals analysis are quite beneficial (24), though the transformer layer was first created for natural language processing tasks (40). To maximize performance for cardiac diagnostics, the transformer's settings are adjusted in the experiment.

It uses the self-attention mechanism to dynamically determine the relative importance of various incoming data segments. At its core, the transformer transforms input vector $X = \{X_1, X_2, \ldots, X_n\}$ into queries ($Q$), keys ($K$), and values ($V$) through weight matrices $W^Q, W^k$ and $W^v$, respectively. The essence of its operation is encapsulated in the attention scores, calculated as

$$Attention(Q, K, V) = softmax\left(\frac{Qk^T}{\sqrt{d_k}}\right) \times V \quad (19)$$

where $d_k$ represents the dimensionality of the key vectors used to scale the dot products. It is expressed as

$$d_k = \frac{d_{model}}{h} \qquad (20)$$

where $d_{model}$ is the embedding dimension, $h$ is the number of attention heads.

To process information across various representation subspaces, The Transformer uses multi-head attention, which is expressed as

$$MultiHead(Q, K, V) = Concat(head_1, \ldots \ldots, head_h)W^o \quad (21)$$

where $W^o$ represents another weight matrix, and each head is expressed as:

$$head_i = Attention(QW_i^Q, KW_i^K, VW_i^v) \qquad (22)$$

A position-wise feed-forward network (FFN) applies two linear transformations to each position identically after attention, with a ReLU activation. It is expressed as

$$FFN(x) = \max(0, xW_1 + b_1)W_2 + b_2 \qquad (23)$$

The experiment's optimized values for the transformer layer are the two heads, thirty-two feed forward, thirty-two channel count, and 20% dropout rate. It is essential to comprehend the temporal dynamics in the data that long-range relationships are effectively captured by the transformer's self-attention mechanism, which sets it apart from conventional recurrent neural network layers. It helps identify intricate patterns and changes that cut across many data segments. Additionally, because of its capacity to analyze and weigh pertinent data across the board, it is well-suited to identify the complex aspects found in heart sounds, helping to create diagnostic models that are more reliable and accurate.

The dropout layer reduces overfitting by arbitrarily deactivating some neurons during training. The experiment has a 20% dropout rate. Max pooling is utilized by downsampling the spatial dimensions of the data and lowering its resolution. This layer chooses the maximum values within predefined regions to extract the most prominent characteristics from the input; the pool size and stride of max pooling are two.

2) **Block 2:** The absence of max pooling and dropout layers sets Block 2 apart from Block 1. Instead, every layer in block -2 has parameters identical to those in block -1. It indicates that no neurons are involuntarily turned off during training, and the data's original resolution and spatial dimensions are preserved.

3) **Decoder:** The global average pooling layer, dropout, fully connected (FC) layer, and softmax layer make up the decoder. The decoder transforms the encoded representation of the input data into an output-helping format. The decoder processes the retrieved features obtained from the last layer. A global average pooling layer pools the temporal sequence and obtains a single value for each vector. After that, a subset of neurons is randomly deactivated during training to minimize overfitting using a 20% dropout rate. Before the softmax layer, a fully connected layer with 32 neurons is used.

### D. Training and Testing

The experiment is intended to assess the performance of heart sound classification and the effective minimum signal required to classify heart sounds accurately. The proposed model is developed and tested using the heart sound signals of 751 people. The dataset is divided into 95% and 5% sections for training and testing. After that, the training dataset is divided into 95% and 5% segments once more to train and validate the model. To prevent model biases, the dataset is divided into patient-wise segments. Since each patient has at least four or more signals gathered from four locations of the chest, all patient signals are preserved in the same splitter dataset throughout the experiment.

The ratio of CHD to non-CHD is 63:37. It demonstrated the imbalance in the dataset. It is also considered when the dataset is divided to preserve the ratio of CHD to non-CHD in the training, validation, and testing datasets. It is imperative to tackle the class imbalance issue in the clinical implementation of deep learning algorithms. The class imbalance problem for heart sound samples from CHD and non-CHD patients was addressed in this study by using distinct class weights in the loss function. The weight of non-CHD samples increases when mispredicted; in other words, the model penalizes classes with erroneous predictions differently. The model favors samples from underrepresented classes since classes with higher weights contribute significantly to the loss. This study used the inverse frequency function to determine the class weights. While training the model, equation (24) provides the weight for each kind of class.

$$W_i = \frac{N}{K \sum_{n=1}^{N} t_{ni}} \qquad (24)$$

In this case, $W_i$ is the class weight of class i, N represents the total number of samples, K is the number of classes, and $t_{ni}$ indicates that sample nth is a member of class $i_{th}$.

### E. Evaluation Parameter

The suggested algorithm's performance is assessed in terms of Accuracy, Sensitivity/Recall, Specificity, Precision, and F1-measure for each experiment. Equations (25 to 29) show these parameters' mathematical expressions. These metrics are frequently employed to assess how well classification algorithms work.

Accuracy is important as it reflects how well the algorithm's predictions align with the ground truth or expected results. It considers both true positive and true negative instances, providing an overall evaluation of the algorithm's performance. Accuracy is expressed in mathematics as

$$Accuracy = \frac{TP + TN}{TP + TN + FP + FN} \qquad (25)$$

Sensitivity, sometimes referred to as recall or true positive rate, is an important performance indicator for deep learning models. It assesses how well the model can identify positive examples among all real positive examples. In the binary classification experiment, sensitivity is computed as the ratio of true positives to the total of false negatives and true positives. The key to sensitivity is its emphasis on reducing false negatives. False negatives in medical diagnosis can have serious repercussions. Sensitivity is defined in mathematics as

$$Sensitivity/Recall = \frac{TP}{TP + FN} \quad (26)$$

The capacity of a model to accurately detect negative instances among all real negatives in a dataset is measured by a statistic called specificity. A high specificity means a low false positive rate, which means the model correctly identifies non-positive cases. Specificity in mathematics is defined as

$$Specificity = \frac{TN}{TN + FP} \quad (27)$$

Precision is a statistic expressing the accuracy of the model's positive predictions. When it comes to medical diagnosis, precision refers to the model's capacity to recognize positive cases accurately without mistakenly classifying an excessive number of negative cases as positive. Confidence in the model's positive predictions is bolstered by a high accuracy score, which suggests a low rate of false positives. The definition of precision in mathematics is as follows -

$$Precision = \frac{TP}{TP + FP} \quad (28)$$

The harmonic mean of precision and recall is the F1-score, often employed in classification problems. Recall evaluates the model's capacity to catch all pertinent occurrences of the positive class, whereas precision assesses the accuracy of positive predictions. These two measures are balanced by the F1-score, which thoroughly evaluates a model's performance. It is particularly pertinent when there is an imbalance in class since it provides a simple number that accounts for both false positives and false negatives. In mathematics, an F-1 score is defined as

$$F1 - score = 2 \times \frac{Precision \times Recall}{Precision + Recall} \quad (29)$$

TP represents the number of true positives—positive cases accurately projected to be positive—and TN represents the number of true negatives — negative cases appropriately anticipated to be negative. Thus, the number of false positives (FP) is the number of negative instances mistakenly anticipated to be positive, and the number of false negatives (FN) is the number of positive cases mistakenly predicted to turn out negative.

## IV. RESULTS

### A. Selection of signal duration

The experiment assesses the signal quality, and the proposed model performance is evaluated by varying the signal duration. The collected signal duration is 15 seconds, and it contains 60000 samples. Since the heartbeat cycle duration is 0.6s – 1s (41), the 15s signal contains 15 to 25 heartbeat cycles. Considering the number of heartbeat cycles, the signal is split into 5s and 3s that contain 20000 and 12000 samples, respectively, and the number of heartbeats is 5 to 8 and 3 to 5, respectively. It follows the number of heartbeat cycles recorded in clinical settings when physicians utilize auscultation to aid in disease diagnosis. The total collected signal for 15 seconds is 3435 signals from 751 patients. After splitting into 5s and 3s signals, the total signals are 10305 and 17175, respectivelys.

### B. Selection of quality assessment indicator threshold

The signal quality assessment indicator RMSSD in the 0.2 - 1 range and ZCR in the 0.2 - 1 range are selected as working range as a threshold value, with 0.1 step size. For 15s and 3s, the signal covered around 99.4% and 99.6%, respectively, within the 0.3 - 0.5 and 0.3 - 4 ranges of the preset indicator threshold RMSSD and ZCR. When ZCR took the minimum threshold value of 0.2 at that time, whatever the RMSSD value, this indicator did not play an influential role in quality assessment. It captured around 18%, 16%, and 20% signals on those thresholds suitable for 15s, 5s and 3s signals, respectively. Notably, 0.5 - 1 value of both RMSSD and ZCR indicator threshold values are also ineffective in assessing the signal quality for 15s, 5s, and 3s signals.

1. **15s signal:** For the 15s signal, the proposed model performed better in classifying the heart sound at ZCR=0.3 than other values of ZCR, while the RMSSD value is in the 0.2 - 1 range shown in Figure 8. Notably, the highest accuracy of heart sound classification is 93.67% at ZCR = 0.3 and RMSSD in the range of 0.4 - 1. The accuracy decreased and remained the same at RMSSD in the 0.4 - 1 range and ZCR in the 0.4 - 1 range. However, the chosen quality assessment indicators threshold to get the best performance of classifying heart sound for 15s signal are the values 0.3 of ZCR and 0.4 of RMSSD. The patient demographics, including age and gender, as displayed in Tables V and VI, are also used to assess the suggested model. With increasing age, the accuracy of the model performs worse. However, women outperform men in terms of model performance.

2. **5s signal:** The best accuracy of classifying heart sounds for the 5s signal at the value 0.2 of ZCR and the RMSSD value 0.2 - 1 range is shown in Figure 9. However, around 16% of signals are considered suitable at those values, which is ineffective. Notably, the highest accuracy of heart sound classification is 93.69% at ZCR = 0.4 and RMSSD in the range of 0.4-1. The model's performance is decreased while increasing the value of ZCR from 0.5 - 1 and the RMSSD value from 0.5 - 1. To consider the minimum value of quality assessment indicators and the highest accuracy to classify the heart sound for 5s signals, the value 0.4 of ZCR and RMSSD is chosen. The patient demographic information, including age and gender, as displayed in Tables V and VI, is also used to assess the suggested model. For 5s signals, the model's classification accuracy of heart sound decreases with age increasing. On the other hand, the model performance is better for females than males.

TABLE V
MODEL PERFORMANCE TO THE AGE RANGE FOR 15S, 5S, AND 3S SIGNALS.

| Duration | Age/Param | Accuracy | Sensitivity/Recall | Specificity | Precision | F1-Score |
|---|---|---|---|---|---|---|
| 15s | 0 to 1 Years (N=174) | 0.9375 | 1.0000 | 0.9167 | 0.8000 | 0.8889 |
| | 1 to 5 Years (N=286) | 0.9559 | 1.0000 | 0.9143 | 0.9167 | 0.9565 |
| | 5 to 10 Years (N=216) | 0.9310 | 0.9302 | 0.9333 | 0.9756 | 0.9524 |
| | More than 10 Years (N=75) | 0.8750 | 1.0000 | 0.7143 | 0.8182 | 0.9000 |
| 5s | 0 to 1 Years (N=174) | 0.9167 | 0.7500 | 0.9722 | 0.9000 | 0.8182 |
| | 1 to 5 Years (N=286) | 0.9700 | 0.9898 | 0.9510 | 0.9510 | 0.9700 |
| | 5 to 10 Years (N=216) | 0.8908 | 0.8760 | 0.9333 | 0.9741 | 0.9224 |
| | More than 10 Years (N=75) | 0.8792 | 0.8630 | 0.9643 | 0.9830 | 0.9191 |
| 3s | 0 to 1 Years (N=174) | 0.9375 | 0.7500 | 1.0000 | 1.0000 | 0.8571 |
| | 1 to 5 Years (N=286) | 0.9619 | 0.9657 | 0.9583 | 0.9548 | 0.9602 |
| | 5 to 10 Years (N=216) | 0.9424 | 0.9419 | 0.9600 | 0.9837 | 0.9073 |
| | More than 10 Years (N=75) | 0.9125 | 0.9556 | 0.8571 | 0.8958 | 0.9247 |

TABLE VI
MODEL PERFORMANCE TO GENDER RANGE FOR 5S AND 3S SIGNALS.

| Duration | Gender/Param | Accuracy | Sensitivity/Recall | Specificity | Precision | F1-Score |
|---|---|---|---|---|---|---|
| 15s | Male (N=447) | 0.9189 | 0.9767 | 0.8387 | 0.8936 | 0.9333 |
| | Female (N=304) | 0.9524 | 0.9565 | 0.9474 | 0.9565 | 0.9565 |
| 5s | Male (N=447) | 0.9054 | 0.8828 | 0.9362 | 0.9496 | 0.9150 |
| | Female (N=304) | 0.9637 | 0.9565 | 0.9727 | 0.9778 | 0.9670 |
| 3s | Male (N=447) | 0.8903 | 0.8651 | 0.9226 | 0.9347 | 0.8986 |
| | Female (N=304) | 0.9516 | 0.9250 | 0.9845 | 0.9867 | 0.9548 |

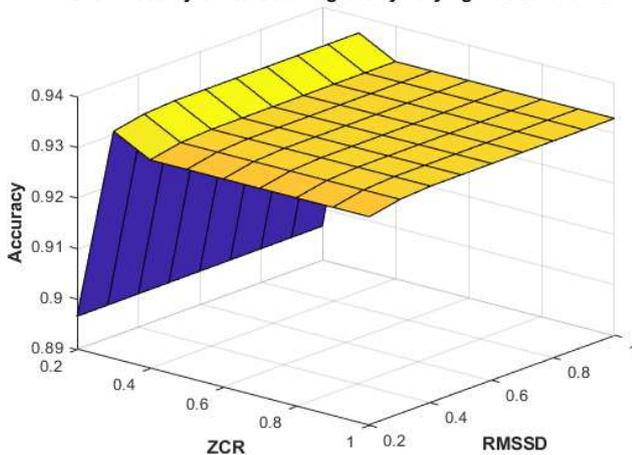

Fig. 8. Based on the RMSSD and ZCR signal quality indicator criteria, this chart shows the accuracy obtained in recognising suitable signals for 15 seconds. Remarkably, the 15-second signal exhibits the maximum accuracy of 93.67% when the RMSSD threshold is set to 0.4 and the ZCR threshold is set to 0.3. The efficacy of these thresholds in signal selection is demonstrated.

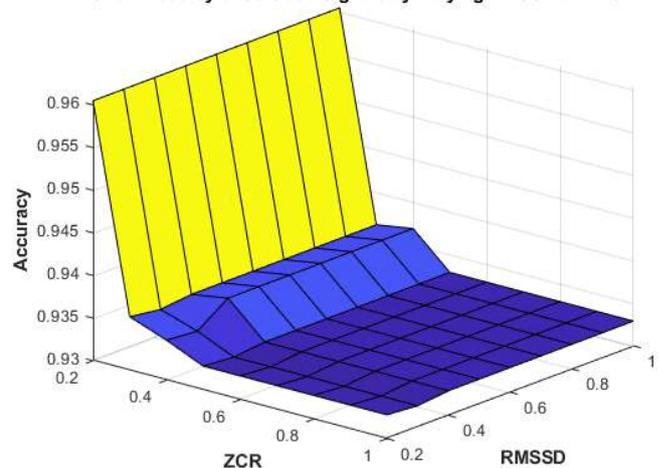

Fig. 9. The accuracy of 5-second suitable signals is shown in this figure. The signals were chosen based on the RMSSD and ZCR signal quality indicator criteria. Remarkably, when both the ZCR and RMSSD thresholds are set to 0.4, the best accuracy for the 5-second signal is 93.69%. It is noteworthy, therefore, that the maximum accuracy of 96.05%, recorded at ZCR = 0.2 in the 0.2 to 1 RMSSD range, is not useful for obtaining appropriate signals.

3. **3s signal:** To evaluate the performance of classifying the heart sound for 3s signals by varying the quality assessment indicators, it is found that the accuracy is increasing by increasing the ZCR from 0.2 - 0.4 while the value of RMSSD is constant within the range of 0.4 – 1 shown in

2Figure 10. The model's performance decreases if ZCR is increasing more than 0.4 while the value of RMSSD is constant within the range of 0.3 - 1. Notably, the best performance of 92.41% for 3s signals is found at the similar quality assessment indicators (RMSSD: 0.4 and ZCR: 0.4) of 5s signals. The patient demographics, including age and gender, as displayed in Tables V and VI, are also used to assess the suggested model. The model's performance in classifying heart sounds decreases as age increases for 3s signals. The performance of the model is better for females than males.

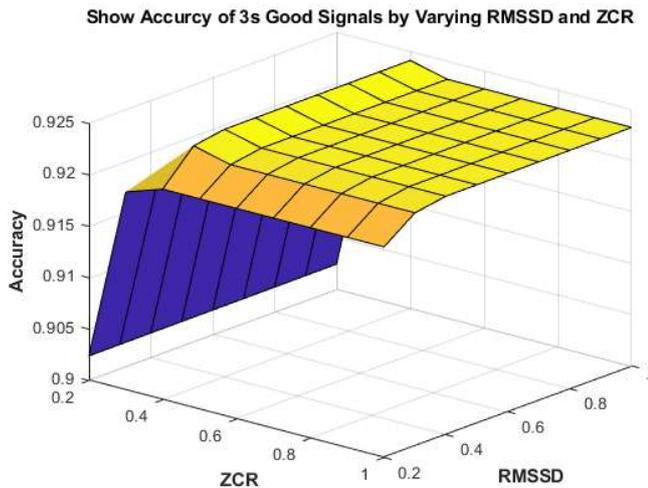

Fig. 10. With respect to the RMSSD and ZCR signal quality indicator criteria, this figure shows the accuracy obtained for 3-second appropriate signals. Notably, the 3-second signal exhibits the best accuracy of 92.41% when both the ZCR and RMSSD thresholds are set to 0.4.

### C. Model Interpretation

To decipher the model's capacity to differentiate between cardiac sound anomalies, activations are shown at every Transformer layer in the proposed network according to the obtained representations. This graphic representation of the algorithm's decision-making process can facilitate understanding how the model interprets and separates CHD specific features. Furthermore, contrasting these visualizations across levels may show hierarchical representations supporting the model's discriminating skills. Visual inspection improved the model's discriminating capability by interacting more deeply with the network. The last two levels of the model, the fifth Transformer layer and global average pooling layer, have the best separation. Starting with the transformer layer, the clustering activations become more efficient using the t-distributed stochastic neighbor embedding (t-SNE) approach (Figure 11). These results imply that the network's deeper layers are essential for improving the model's capacity for heart sound discrimination. It is clear from the t-SNE analysis that as the model moves through the layers, it picks up more distinct and separable characteristics.

## V. DISCUSSION

### A. Duration of the signal's impact on classification performance

This study investigates the effect of heart sound signal length on the heart sound classification performance by varying the signal quality assessment indicators value within the range of 0.2 - 1 with a step size of 0.1. It is critical to analyze the signal quality in pediatric patients' heart sounds—the influence of the signal quality evaluation indicator on the classification performance. The signal quality assessment indicator RMSSD threshold value of 0.4 is the same to get the best accuracy of classifying the heart sound for 15s, 5s, and 3s signals. However, the ZCR threshold value 0.3 is for 15s signals, and 0.4 is for 5s and 3s signals. It indicates that the ZCR value should be selected lower to assess the quality of longer signal duration and higher for the shorter signal duration. The heart sound duration is sensitive to the signal quality assessment indicator threshold ZCR, but the threshold RMSSD is less sensitive. The performances of the model for 15s and 5s are slightly different, though the highest accuracy, 93.69%, is found for the 5s signal.

Furthermore, a typical resting heartbeat is between 60 and 100 beats per minute. It indicates that around 20, 7, and 4 whole cardiac cycles are included in the 15s, 5s, and 3s, respectively. Even though a longer heart sound from a pediatric child has more cardiac cycles, it can occasionally be challenging to obtain a longer heart sound due to motion artifacts and ambient noise. However, shorter heart sounds, such as 3s signals, may be lack the information to be successfully classified. A comparison is made between the proposed method's signal duration, collection area, age range, and classification methods and those of the previously developed heart sound classification model, shown in Table I, which revealed that some of the works that have already been done had used self-collected PCG signal data. In contrast, others have used the Physionet 2016 dataset, which is divided into only healthy and unhealthy classes.

### B. Performance of the demographic data

This study also analyses the proposed model performance on the patient's demographic data. Regarding gender, the model's performance in female patients is better than in male patients for 15s, 5s, and 3s signals, as shown in Table VI. It is because of the different chest depths in gender. As auscultate the heart, the depth of the chest is a crucial consideration. The average chest depth of men is greater than that of women, according to research published in the Oxford Academic Journal (42). The model's performance decreases as age increases for all the selected signal durations in the experiment shown in Table V. The experiment demonstrates that the model's performance falls for all specified signal durations as age increases. It implies that the age range is an essential element influencing the model's performance. More research might be needed to investigate the precise causes of this decline in performance with aging.

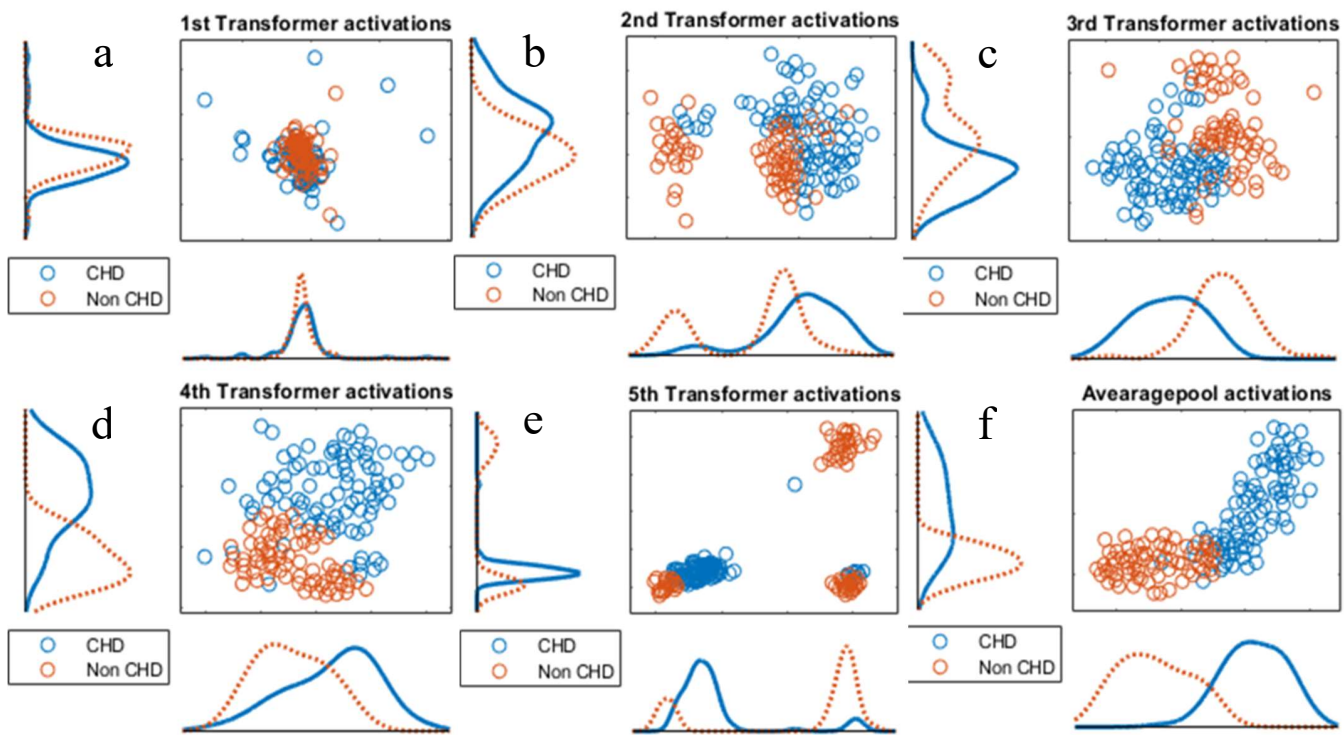

Fig. 11. Analyzing network activations layers to determine how well the model identifies CHD. The figure depicts the internal activations of transformer layers that are grouped using the t-distributed stochastic neighbor embedding (t-SNE) method. This allows for a visual evaluation of the model's ability to discern the existence of CHD. Specifically, (a) through (e) correspond to the activations of the 1st to 5th transformer layers, respectively, while (f) represents the activation at the global average pooling layer.

## V. Conclusion

The study's main objective is to find the minimum signal duration to accurately classify the heart sound and find the optimum signal quality assessment indicator RMSSD and ZCR threshold value. The proposed model is a transformer-based residual one-dimensional convolutional neural network. To improve signal feature extraction, the model efficiently collects temporal information, long-range correlations, and complex patterns across several data segments. Furthermore, recognizing patterns associated with various cardiac diseases aids in creating more accurate and contextually aware classifications. The study showed that the model performed well for 5s signal duration while the signal quality assessment indicators RMSSD and ZCR value 0.4. Finding the minimum signal duration is also suitable for considering developed IoT-based medical devices that can accurately classify the heart sound signal with low-resource hardware — implementing the classification algorithm in a device to facilitate medical service in rural areas.

## Conflict of Interest

Each author has declared that they have no relationships to disclose relevant to this paper's information.